\documentclass[smus]{snow2e}
\usepackage{epsf}
\usepackage{dcolumn}

\newcolumntype{d}[1]{D{.}{.}{#1}}

\newcommand{\hpp}{\mbox{$\Delta^{++}$}}
\newcommand{\hmm}{\mbox{$\Delta^{--}$}}

\def\MPL #1 #2 #3 {{\rm Mod.~Phys.~Lett.}~{\bf#1}\ (#3) #2}
\def\NPB #1 #2 #3 {{\rm Nucl.~Phys.}~{\bf#1}\ (#3) #2}
\def\PLB #1 #2 #3 {{\rm Phys.~Lett.}~{\bf#1}\ (#3) #2}
\def\PR #1 #2 #3 {{\rm Phys.~Rep.}~{\bf#1}\ (#3) #2}
\def\PRD #1 #2 #3 {{\rm Phys.~Rev.}~{\bf#1} \ (#3) #2}
\def\PRL #1 #2 #3 {{\rm Phys.~Rev.~Lett.}~{\bf#1}\ (#3) #2}
\def\RMP #1 #2 #3 {{\rm Rev.~Mod.~Phys.}~{\bf#1}\ (#3) #2}
\def\ZP #1 #2 #3 {{\rm Z.~Phys.}~{\bf#1}\ (#3) #2}
\def\IJMP #1 #2 #3 {{\rm Int.~J.~Mod.~Phys.}~{\bf#1}\ (#3) #2}

\def\thetaw{\theta_W}
\def\gamz{\Gamma_Z}
\def\xw{x_W}
\def\yw{y_W}

\def\tev33{Tev$33$}

\def\rts{\sqrt s}

\def\epem{e^+e^-}

\def\lsim{\mathrel{\raise.3ex\hbox{$<$\kern-.75em\lower1ex\hbox{$\sim$}}}}
\def\gsim{\mathrel{\raise.3ex\hbox{$>$\kern-.75em\lower1ex\hbox{$\sim$}}}}
\def\@versim#1#2{\vcenter{\offinterlineskip
        \ialign{$\m@th#1\hfil##\hfil$\crcr#2\crcr\sim\crcr } }}
\def\gamstar{\gamma^\star}
\def\zstar{Z^\star}

\def\ie{{\it i.e.}}

\def\gam{\gamma}

\def\anti{\overline}

\def\fbi{~{\rm fb}^{-1}}
\def\fb{~{\rm fb}}

\def\gev{\,{\rm GeV}}
\def\tev{\,{\rm TeV}}

\def\rta{\rightarrow}
\def\to{\rightarrow}

\def\hm{H^-}

\def\mz{m_Z}
\def\mw{m_W}

\def\wm{W^-}

\def\emem{e^-e^-}
\def\em{e^-}
\def\ep{e^+}
\def\hmm{\Delta^{--}}
\def\mhmm{m_{\hmm}}

\def\hpp{\Delta^{++}}

\def\hm{\Delta^{-}}
\def\mhm{m_{\hm}}
\def\hzero{\Delta^0}

\def\dpp{\Delta^{++}}
\def\dmm{\Delta^{--}}
\def\electron{$e^-e^-$}

\newcommand{\et}{\mbox{$E_T$}}

\newcommand{\met}{\mbox{$\not\!\!E_T$}}

\newcommand{\gevc}{\mbox{${\rm GeV}/c$}}

\newcommand{\invfb}{\mbox{${\rm fb}^{-1}$}}

\begin{document}
\title{
{\normalsize 
             \hspace*{\fill} UCD-96-30  \\
             \hspace*{\fill} Fermilab-Conf-96/347 \\
             \hspace*{\fill} October, 1996 \\
             \hspace*{\fill}  \\
}
Searching for Doubly-Charged Higgs Bosons at
 Future Colliders\thanks{To appear in ``Proceedings
of the 1996 DPF/DPB Summer Study on New Directions
for High Energy Physics''.  
Work supported by U.S. 
Department of Energy and the National Science Foundation.}}


\author{J.F. Gunion\\ 
{\it Department of Physics, University
of California at Davis, Davis, California 95616} \and
C. Loomis\\ 
{\it Rutgers University, Piscataway, New Jersey 08855} \and
K.T. Pitts \\ 
{\it Fermi National Accelerator Laboratory, Batavia,
Illinois, 60510}}

\maketitle

\thispagestyle{empty}\pagestyle{plain}

\begin{abstract} 
Doubly-charged Higgs bosons 
($\dmm/\dpp$)
appear in several
extensions to the Standard Model and can be relatively light.  
We review the theoretical motivation for these states
and present a study of the discovery reach
in future runs of the Fermilab Tevatron 
for pair-produced doubly-charged Higgs bosons
decaying to like-sign lepton pairs.
We also comment on the discovery potential at
other future colliders.
\end{abstract}

\section{Introduction}
Doubly-charged Higgs bosons ($\dmm$) appear in exotic
Higgs representations such as found in left-right symmetric models.
The current experimental bound is $\mhmm > 45 \gev$
\cite{opal} from a search for $Z^0\to \dmm\dpp$ at LEP.

At the Tevatron, the two production mechanisms with potentially large
cross section are
pair production, $p\overline{p}\to \gamma/Z^0X
\to \dmm\dpp X$ or single production
via $WW$ fusion, $p\overline{p}\to \wm\wm X
\to \dmm X$.  However, existing phenomenological
and theoretical constraints are only easily satisfied if the
$\wm\wm\to\dmm$ coupling is vanishing (or very small).   
Therefore, in this
analysis we will consider the discovery reach for detecting
$\dmm\dpp$ pair production at the Tevatron.

In many models, it is possible for the $\dmm$ to couple to like-sign
lepton pairs, $\ell^-\ell^-$. If
the $\wm\wm\to\dmm$ coupling is vanishing, it is then very
likely that the doubly-charged Higgs will decay to $\ell^-\ell^-$
via the lepton-number-violating coupling.
We will therefore concentrate upon $\dmm\to e^-e^-$, 
$\dmm\to \mu^-\mu^-$ and
$\dmm\to \tau^-\tau^-$.

Alternatively, if the $\dmm\to\ell^-\ell^-$ and $\dmm\to\wm\wm$ couplings
are both vanishing or very small, then the $\dmm$
can have a sufficiently long lifetime that it will decay outside the
detector. Identification of the $\dmm\dpp$ pair via the associated $dE/dx$
distributions in the tracking chamber would then be possible.

\section{Theoretical Motivation}
  
Doubly-charged scalar particles abound in exotic
Higgs representations and appear in many models \cite{hhg,lr89,triplets}.
For example, a Higgs doublet representation with
$Y=-3$ contains a doubly-charged $\hmm$ and a singly-charged $\hm$.
If part of a multiplet with a neutral member, a $\hmm$
would immediately signal
the presence of a Higgs representation with total isospin
$T=1$ or higher. 
Most popular are the complex $Y=-2$ triplet Higgs representations,
such as those required in left-right symmetric models, 
that contain a $\hmm$, a $\hm$ and a $\hzero$.

In assessing the attractiveness of a Higgs sector
model containing a $\hmm$ many constraints need to be considered.
For triplet and higher representations containing a neutral
member, limits on the latter's vacuum expectation value (vev)
required for $\rho\equiv \mw^2/[\cos^2\thetaw\mz^2]=1$ at tree-level are 
generally severe.
(The first single representation beyond $T=1/2$ for which $\rho=1$
regardless of the vev is 
$T=3,Y=-4$, whose $T_3=0$ member is doubly-charged.)
Models with $T=1$ and $T=2$ can
have $\rho=1$ at tree-level by combining representations.
However, such models generally
require fine-tuning in order to preserve $\rho=1$ at one-loop.
The simplest way to avoid all $\rho$ problems is to either consider
representations that simply do not have a neutral member
(for example, a $Y=-3$ doublet or a $Y=-4$ triplet representation), or else
models in which the vev of the neutral member
is precisely zero. We will only consider models of this type in what follows.

Further constraints on Higgs representations arise
if we require unification of the coupling constants
without intermediate scale physics. In the Standard Model, 
unification is possible for a relatively simple Higgs sector that includes 
a single $|Y|=2$ triplet in combination
with either one or two $|Y|=1$ doublets (the preferred number of
doublets depends upon the precise
value of $\alpha_s(\mz)$). 
In the case of the minimal supersymmetric extension of the
Standard Model, precise unification requires exactly two doublet
Higgs representations (plus possible singlet representations); 
any extra doublet representations (including
ones with a doubly-charged boson) or any number
of triplet or higher representations would destroy unification. 
However, by 
going beyond the minimal model and including appropriate
intermediate-scale physics, supersymmetric models (in particular,
supersymmetric left-right symmetric models \cite{lrunification}) with triplet
and higher representations can be made consistent with unification.

In short, the popular two-doublet MSSM need not be nature's choice.
We should be on the look-out for signatures of exotic Higgs
representations, the clearest of which would be the
existence of a doubly-charged Higgs boson.
Thus, it is important to consider how
to search for and study such a particle.  

The phenomenology of the $\hmm$ derives from its couplings.
Tri-linear couplings of the type $\wm\wm\rta\hmm$ 
are not present in the absence of an enabling non-zero vev
for the neutral member (if present) of the representation,
and $q^\prime \anti q \hmm$ couplings are obviously absent.
There are always couplings of the form $Z,\gamma\rta \hmm\hpp$.
In addition, and of particular interest,
there is the possibility of lepton-number-violating
$\ell^-\ell^- \rta\hmm$ couplings in some models.
For $Q=T_3+{Y\over 2}=-2$ the allowed cases are:
\begin{equation}
\begin{array}{l}
 \ell^-_R\ell^-_R\rta \hmm(T=0,T_3=0,Y=-4) \,, \\ 
 \ell^-_L\ell^-_R\rta \hmm(T={1\over 2},T_3=-{1\over 2},Y=-3) \,,\\  
 \ell^-_L\ell^-_L\rta \hmm(T=1,T_3=-1,Y=-2)\,.
\end{array}
\label{helicitycases}
\end{equation}
Note that the above cases do not include the $T=3,Y=-4$ representation 
that yields $\rho=1$, nor the $T=1,Y=-4$ triplet with no neutral
member, but do include the $T=1/2,Y=-3$ doublet
representation with no neutral member, 
and the popular $T=1,Y=-2$ triplet representation.
In left-right symmetric models
there is a `right-handed' and a `left-handed' Higgs triplet,
both with $|Y|=2$. Our analysis applies to the left-handed triplet
(whose neutral member must have a very small vev to preserve $\rho=1$);
the phenomenology of the right-handed triplet is completely different.

In the case of a $|Y|=2$ triplet representation (to which we now
specialize) the lepton-number-violating coupling to (left-handed)
leptons is specified by the Lagrangian form:
\begin{equation}
{\cal L}_Y=ih_{ij}\psi^T_{iL} C\tau_2\Delta\psi_{jL}+{\rm h.c.}
\,,
\label{couplingdef}
\end{equation}
where $i,j=e,\mu,\tau$ are generation indices, 
the $\psi$'s are the two-component
left-handed lepton fields ($\psi_{\ell L}=\pmatrix{\nu_\ell,\ell^-}_L$), and
$\Delta$ is the $2\times 2$ matrix of Higgs fields:
\begin{equation}
\Delta=\pmatrix{\hm/\sqrt{2} & \hmm \cr \hzero & -\hm/\sqrt{2} \cr}\,.
\end{equation}


Limits on the $h_{ij}$ coupling strengths come from many sources. 
Experiments that place limits on the $h_{ij}$ by virtue of the $\hmm\rta
\ell^-\ell^-$ couplings include Bhabha scattering, $(g-2)_\mu$, 
muonium-antimuonium conversion, and $\mu^-\rta e^- e^- e^+$. 
These limits \cite{lr89,hmpreprint} suggest small off-diagonal
couplings (as assumed in our analysis). Writing
\begin{equation}
|h_{\ell\ell}|^2\equiv c_{\ell\ell} \mhmm^2(\gev)\,,
\label{hlimitform}
\end{equation}
the limits imply $c_{ee}\lsim 10^{-5}$ and $c_{\mu\mu}\lsim 6\cdot 10^{-5}$.

Regarding production mechanisms,
the fusion process \cite{lr89,triplets,huituetal}, 
$\wm\wm\rta \hmm$, is absent since the required
tri-linear coupling is zero if the vev of the neutral member (if
there is one) of the Higgs representation is zero (as we assume
so that $\rho=1$ naturally). Single production of $\hmm,\hpp$ ($\hmm$)
is possible in $\epem$  ($ep$) collisions at LEP2 (HERA) via diagrams
involving the $\hmm\to\em\em$ or $\hpp\to \ep\ep$ couplings.
If $c_{ee}$ saturates its upper limit, then LEP2
and HERA will probe up to $\mhmm\sim 150\gev$ \cite{lpet,apet}.
However, it is likely that $c_{ee}$ is much smaller
than its current bound and that these sources of single production
will be negligible.

Thus, we focus on $\gamstar,\zstar\to\dmm\dpp$ pair production, the cross
section for which is determined entirely by the quantum numbers
of the $\dmm$. For a general spin-0 boson $B$, 
with weak isospin $T_3$ and charge
$Q$, and a fermion $f$, with $t_3$ and $q$, the $f\anti f\to B\anti B$
pair-production cross section is:
\begin{eqnarray}
\sigma^{\rm pair}(s)&=\left( {\pi\alpha^2 \beta^3 s\over 6}\right)  &
\Biggl\{ 2 Q^2q^2 P_{\gam\gam}+P_{\gam Z}{2Qq A(a_L+a_R)\over \xw\yw}
\nonumber\\
&&+P_{ZZ}{A^2(a_L^2+a_R^2)\over \xw^2\yw^2}\Biggr\}
\,,
\end{eqnarray}
where $s$ is the $f\anti f$ center of mass energy squared,
$\beta=\sqrt{1-4m_B^2/s}$,
$\xw=\sin^2\thetaw$, $\yw=1-\xw$,
$A=T_3-\xw Q$, $a_L=t_3-\xw q$, $a_R=-\xw q$, $P_{\gam\gam}=s^{-2}$,
$P_{ZZ}=[(s-\mz^2)^2+\mz^2\gamz^2]^{-1}$, and 
$P_{\gam Z}=(s-\mz^2)P_{ZZ}/s$. We will consider a
$\dmm$ with $T_3=-1,Q=-2$. An extra factor of 1/3 is required
for color averaging in $q\anti q$ annihilation in $pp$ or $p\bar p$
collisions.

\begin{figure}[ht]
\leavevmode
\epsfxsize=3.250in
\hspace*{0.25in}
\epsffile{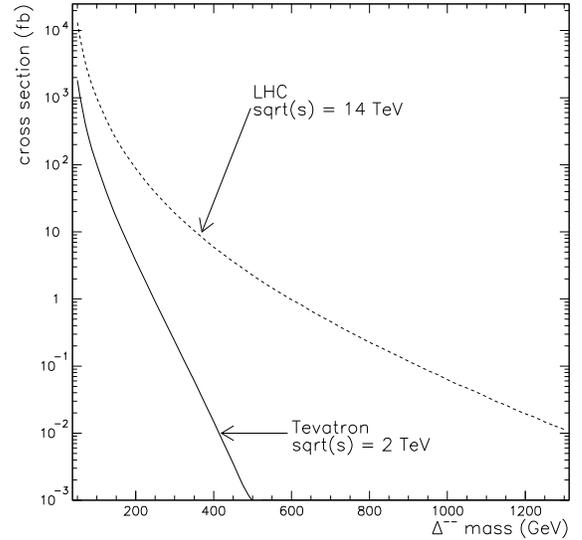}
\caption{$\dpp/\dmm$ pair production
cross section as a function of $\dmm$ mass
for both the Tevatron and the LHC.}
\label{fig:xsec}
\end{figure}

In $\epem\to\dmm\dpp$, kinematic reach is limited to $\mhmm\lsim \rts/2$,
\ie\ no more than about $230-240\gev$ at a future $\rts=500\gev$ NLC.
We will find that the discovery reach at the Tevatron can cover much, if not
all, of this range, depending upon the dominant $\hmm$ decay mode.
The mass reach for pair production at a $pp$ collider increases
rapidly with machine energy.  
Figure~\ref{fig:xsec} shows the $\dmm\dpp$
pair production cross section for both the Tevatron (at $\rts=2\tev$)
and the LHC. At the Tevatron, $\sigma^{\rm pair}\sim 0.9(0.24)\fb$
at $\mhmm=250(300)\gev$. With
total accumulated luminosity of $30\fbi$ (as planned
for the \tev33\ upgrade)
there would be about 27(7) $\hmm\hpp$ events.  The marginality
of the latter number makes it clear that $\mhmm\lsim 300\gev$ will be
the ultimate mass reach possible at the Tevatron.

Decays of a $\hmm$ are generally quite exotic \cite{lr89,triplets}.
For $\sim 0$ $\hmm\rta\wm\wm$ coupling, the only two-body 
decays that might be important are
$\hmm\rta \hm\wm$, $\hmm\rta\hm\hm$ and, if the lepton coupling
is present, $\hmm\rta \ell^-\ell^-$. Typically, the $\hmm$ and $\hm$
have similar masses, in which case
$\hmm\rta\hm\hm$ is likely to be disallowed.  Thus,
we will focus on the $\hm\wm$ and $\ell^-\ell^-$ final states.
For a $T=1,Y=-2$ triplet we find \cite{lr89,triplets}
\begin{equation}
\begin{array}{l}
\Gamma_{\hmm}^{\hm\wm}= {g^2\over 16\pi} {m_{\hmm}^3\beta^3
\over\mw^2}\sim (1.3\gev) \left({\mhmm\over
100\gev}\right)^3\beta^3\,, \\
\Gamma_{\hmm}^{\ell^-\ell^-}=
{|h_{\ell\ell}|^2\over 8\pi} m_{\hmm}
\sim (0.4\gev) \left( {c_{\ell\ell}\over 10^{-5}}\right)
\left({\mhmm \over 100\gev}\right)^3\,.
\end{array}
\label{widtheq}
\end{equation}
where $\beta$ is the usual phase space suppression factor,
and we used Eq.~(\ref{hlimitform}).
For example \cite{lr89}, if
$\mhmm=360\gev$, $\mhm=250\gev$
we find $\Gamma(\hmm\rta\hm\wm)\sim 2\gev$ and $\Gamma(\hmm\rta\ell^-\ell^-)=
19\gev \left({c_{\ell\ell}\over 10^{-5}}\right)$.
If any $c_{\ell\ell}$ is near $10^{-5}$
then $\Gamma_{\hmm}^{\ell^-\ell^-}> \Gamma_{\hmm}^{\hm\wm}$ is likely.
Since there are currently no limits on $c_{\tau\tau}$, the $\tau^-\tau^-$
channel could easily have the largest partial width and be the dominant
decay of the $\hmm$. On the other
hand, if all the $c_{\ell\ell}$  are very small
then the $\hm\wm$ mode is quite likely to be dominant if it
is kinematically allowed. 
The implications for detection of $\hmm\hpp$ pairs will now be discussed.


\section{Simulation and Reconstruction}

The signal and backgrounds are simulated
with the PYTHIA Monte Carlo, which has been
modified to allow the process:
\begin{equation}
p\overline{p}\to Z^0/\gamma X
\to \dmm\dpp X,
\end{equation}
with the $\dmm$ then forced to decay
to like-sign lepton pairs.
 
The events are then fed to a CDF detector
simulation which includes the geometry of 
the Run I CDF detector.  For the Main Injector
runs of the Tevatron, the CDF and D0
detectors will both be upgraded to handle higher
instantaneous luminosity.
In addition, the acceptances of the upgraded
detectors will improve.
This simulation includes muon 
coverage for $|\eta| < 1$, which will be improved
to $|\eta| < 1.5$ for Run II.  This results in 
approximately a $20\%$ improvement in acceptance for
this process.

Events are passed through the normal CDF event
reconstruction package.  Muon candidates must have
tracks in both the central tracking and the muon
chambers, electron candidates must have a track and
an isolated electromagnetic calorimeter cluster.
The lepton momenta are 
determined from the central tracking chamber
and, if fiducial, the silicon microvertex detector.  
For tracks which do not pass through the microvertex
detector, the fit is performed assuming that the 
track originated from the interaction point.  This
so-called ``beam-constraint'' significantly improves
momentum and, hence, mass resolution. 

\section{$\dmm\to$\ \electron,
$\dmm\to \mu^-\mu^-$}

For the case where the $\Delta$ decays to like-sign 
leptons (excluding taus), the signature
is a spectacular $4e$ or $4\mu$ final state.
Here we will focus upon the $4\mu$ final
state.  Backgrounds are very similar for the two channels,
although the discovery reach will be slightly higher
in the electron channel due to better mass resolution
and larger electromagnetic
calorimeter coverage.

The dominant backgrounds in the $4\mu$ mode 
(accepting  at least 2 same-sign $\mu$'s
as described below) 
arise from electroweak processes
where real high-$p_T$ \  muons are created from $W$ or $Z$
decays along with either fake muons or muons from 
heavy flavor decay.  The backgrounds are 
diboson production ($ZZ\rta 4\mu$,
$WZ\rta 3\mu+\nu$, $WW\rta 2\mu+2\nu$); 
$t\overline{t}$ production 
 ($t\overline{t} \to \mu^+ \nu b\  \mu^- 
\overline{\nu} \overline{b}$); 
and
boson plus jets 
($W + jets$, $Z + jets$),
where $W \rta \mu \nu$, $Z \rta \mu^+\mu^-$
and the jets produce real or fake muons.
We use the measured cross sections for $t\overline{t}$, $W+jets$ and
$Z+jets$ \cite{tt,wjet,zjet} and the calculated cross sections
for $WZ$ and $ZZ$ production \cite{wwwz}.  The PDF world average
branching
ratios are used
for $Z\to \mu^{+} \mu^{-}$ (0.03367) and 
$W\to \mu^{+}\nu_\mu$ (0.104) \cite{pdg}.


\begin{figure}[ht]
\leavevmode
\epsfxsize=3.25in
\hspace*{0.25in}
\epsffile{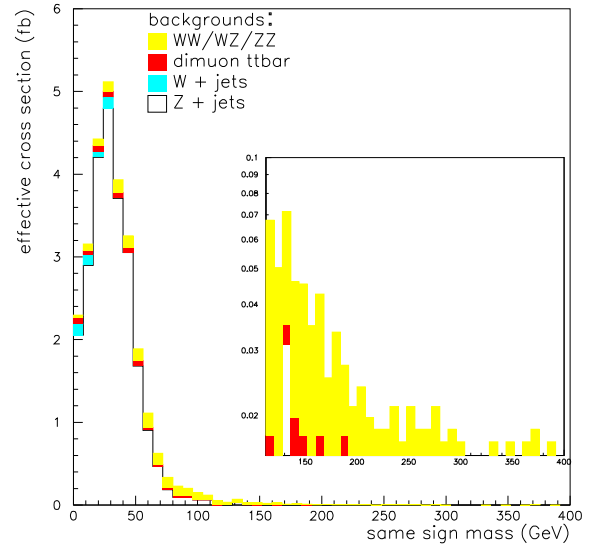}
\caption{Background contributions to the same sign mass plot
after all cuts.  As can be seen in the inset, the dominant
background above $100\gev$ is from diboson production.
The $\rm{N}_{jet}\le 1$ jet requirement 
removes most of the $t\overline{t}$ 
background.}
\label{fig:bk}
\end{figure}

\begin{figure}[ht]
\leavevmode
\epsfxsize=3.250in
\hspace*{0.25in}
\epsffile{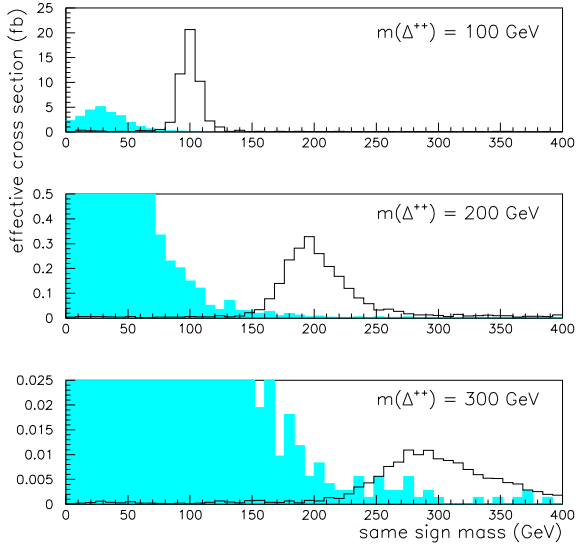}
\caption{The effective cross section
(true cross section $\times$ efficiency $\times$
acceptance) for signal (unshaded) and 
background (shaded) for $\dmm$ masses of
$100$, $200$ and $300\gev$ respectively, 
after all cuts have been applied.}
\label{fig:mass}
\end{figure}

With a small event rate and small backgrounds, it is
desirable to keep the selection criteria as loose
as possible.  To simulate the trigger, we require one
central muon with $p_T > 12 ~\gevc$.   Next, at least
two same-sign muons are required.   This is sufficient
to easily isolate a signal in the low mass region
$m_{\mu\mu} < 100 \gev$, but the background from 
dimuon top decay becomes an issue at higher
masses.   We expect very little additional activity
in $\Delta$ pair production other than the energy
recoiling against the virtual $Z/\gamma$.  For dimuon
top decay backgrounds, there is additional jet activity
from the two $b$ decays.  The third muon is supplied by one
of these $b$ decays.  The background in the high mass region
from top decays can be greatly suppressed by requiring
that the event has no more than one 
jet seen in the calorimeter 
($|\eta|<2.5$) with more than $7.5\, \gev$ of transverse
energy ($E_T$).

Figure~\ref{fig:bk} shows the same-sign muon mass
contributions from each
of the backgrounds listed above after all cuts.  
Above $100\gev$ in same-sign mass, the dominant
background is diboson production.

Figure~\ref{fig:mass} shows the same-sign dimuon 
invariant mass distributions for both signal and background
for three different $\Delta$ masses: 100, 200 and $300\gev$
after all cuts have been applied.  
The signal-to-background ratio remains high at 
$\mhmm = 300\gev$, although the dimuon mass resolution
is worsening.   The dimuon mass resolution is dominated by
the $p_T$ resolution of the detector, which worsens at
higher momenta.  For the case 
$\dmm\to e^-e^-$ the dielectron mass resolution 
does not degrade as rapidly due to the energy
resolution of the electromagnetic calorimeter.
In the case of a high mass search for $\dmm$
decaying
to muon or electron pairs, the technique would be a counting
experiment, looking for an excess of high mass pairs
over the small background. 

The simulated data shown in Figures~\ref{fig:bk} and 
\ref{fig:mass} represent the response of the Run I CDF
detector.
The product of the efficiency and acceptance for
a signal event to produce at least one entry in
the same-sign mass plot depends upon $\mhmm$, but is
typically $50\%$.  If we assume
that the efficiency will be the same for the
Run II detector, 
scaling the acceptance to the improved muon
coverage ($|\eta|<1.5$) brings this number
up to $60\%$, an improvement factor 
in the accepted signal
of $1.2$.   The corresponding scaling of
the acceptance for the background is found to
be $\sim\! 2$ over the entire 
mass region above $50\gev$.

In addition to the significance of a bump in the same-sign
mass distribution, there is additional information in the 
number of high-$p_T$ muons in the event.   With the Run I
CDF detector used in this simulation, approximately
$20\%$ of the signal events
have four found muons, yielding
two entries in the mass distribution, while $\sim\! 6\%$
of the background events have more than three muons such
that both same-sign pairs have mass greater than $50$\gev.  
For the Run II detector, these numbers  go up to
approximately $40\%$ for signal and $11\%$ for 
the background.
The probability
that a background event would have four muons \bf and \rm 
both same-sign combinations near one-another in mass is 
exceedingly small.   We therefore conclude that production
of enough events so that two events are measured to have
four muons (in addition to the other same-sign dimuon mass
entries from $2$ and $3$ muon events) 
will be more than adequate to establish a signal for
the $\dmm$.  
As an example using the numbers above: if $10$ events are
produced, $6$ events would produce at least one same-sign
mass pair.  Of those $6$ events, $2$ (from 2.4) 
would have four found
muons and $4$ (from 3.6) 
would have $2$ or $3$ found muons, yielding
$8$ entries in the same-sign mass plot on a background of
approximately one same-sign dimuon mass pair and zero
four muon events.  
Tri-muon events offer little additional evidence for
$\dmm$ production, since background events often have
two real, opposite-signed leptons in addition to one lepton
from either a fake or heavy flavor decay.


Using the criteria that $10$ pair-produced events would
lead to an unambiguous discovery of the $\dmm$, we conclude
that a reach of approximately $200$, $250$, 
$300 ~\rm GeV$ in the mass of
the $\dmm$ could be achieved in $2$, $10$, $30\fbi$
of Tevatron running, respectively, for the cases where
$\dmm \rightarrow e^-e^-$ and 
$\dmm \rightarrow \mu^-\mu^-$.  


\section{$\hmm\to\tau^-\tau^-$}

\noindent Unlike the electron and muon channels, reconstructing an
invariant mass in the tau channel is problematic because of the
neutrinos involved in their decay.  We therefore use a counting method
to estimate the reach of a doubly charged Higgs search for Run~II.

Tau lepton identification is not trivial at a hadron collider.
Identification efficiencies are much lower than for electrons or muons
($\sim 50\%$) and fake rates from QCD jets are significant ($\sim
0.5\%$).  Nonetheless, searching the tau lepton channel is worthwhile
because the doubly charged Higgs may preferentially couple to the taus,
and the tau lepton offers the possibility of measuring the spin of its parent.

Selection of tau lepton candidates  which decay into hadrons 
is detailed in  \cite{tau}.
The algorithm begins by looking for  jets
in the calorimeter.
The tau candidate must have one
or three charged particles in a $10^\circ$ cone about the jet axis and
no additional charged particles in a cone of $30^\circ$.  In addition,
the tau candidate must have the correct total charge ($\pm 1$)
and have a mass
consistent with a tau lepton.  At a minimum, the cluster must have
$E_T>10$~\gev\ and the largest $p_T$ of an associated charged particle
must exceed $10$~\gevc.  Finally, the tau candidate cannot be consistent
with an electron.

Currently, a fiducial cut of 
$|\eta|<1$ 
is required to maintain good charged particle
tracking efficiency.  The tracking coverage 
for the Run~II detector will
be significantly larger and tracks will be measured
with improved resolution.   
The
corresponding gains in tau lepton 
acceptance have not been included in
the results below.  

Three types of triggers are considered
for this search:  inclusive electron and muon triggers for the case
where at least one of the taus has decayed leptonically, an \met\
trigger which relies on the neutrinos in the tau decay, and a
dedicated tau lepton trigger which identifies tau leptons from
tracking or calorimetry information early in the triggering system.
These triggers 
correspond to triggers
used in Run~I and allow the background from fake taus to be estimated
from data.
For the estimates below the logical OR of the missing \et\ and the
inclusive lepton triggers has been used.  The threshold on the
\met\ trigger is $35$~\gev\ and the threshold on the inclusive lepton
triggers is $\et>20$~\gev.  The primary tau lepton in an event must
have $\et>20$~\gev.  

Pair production of doubly charged 
Higgs produces events with four taus
in the final state.  Approximately $60\%$ of these events contain three
or more taus which decay into hadrons.  To reduce the background from
fake taus to a reasonable level, a selected event must have at least
three identified hadronic tau lepton candidates
and an
additional jet, electron, muon, or hadronic tau candidate.  
The expected number of signal
events in a $10$~\invfb\ sample for various $\hmm$  masses is
shown in Table~\ref{exptt}.

\begin{table}
\caption{Expected $\hmm\to\tau\tau$ events passing all
cuts in $10$~\invfb\label{exptt}.}
\begin{center}
\begin{tabular}{cd{2}@{$\pm$}d{2}}
\hline
\hline
{$M_{\hmm}$ (\gev)} &
\multicolumn{2}{c}{events} \\
\hline
$ 50 $ & 19.    &  4.    \\
$100 $ &  8.8   & 0.6  \\
$150 $ &  3.07 & 0.20  \\
$200 $ &  0.72 &  0.11  \\
$250 $ &  0.23  & 0.03  \\
\hline
\hline
\end{tabular}
\end{center}
\end{table}

The backgrounds are expected to come from two sources---processes
which produce real taus and processes which produce jets which
fluctuate to imitate tau leptons.  The expected number of events
containing multiple fake tau leptons was estimated from the Run~I data
sample.  No events pass the selection requirements.

Top and diboson production are expected to be the largest source of
background events with real tau leptons.  Both of these were
estimated from Monte Carlo.
Again, no events passed the strict
topology cuts.  Given the large data samples expected in Run~II, both
of these backgrounds should be measured rather precisely.

Given that no background events pass the selection requirement, it is
difficult to quantitatively define the number of events necessary to
claim a discovery.  Nonetheless, the background is likely to be quite
small, so a handful of events should be considered significant.
Arbitrarily taking five events as the standard, a search for doubly
charged Higgs in the tau channel would have a reach of approximately
$130$ and $180$~\gev\ in samples of $10$ and $30$~\invfb, 
respectively.  The
cuts, trigger, {\it etc.}\ have not been optimized, so these should
be considered conservative estimates of the Run~II reach.






\section{Conclusions}

Although currently out of favor because of the success of the
minimal supersymmetric model, there are well-motivated models
containing triplet and other Higgs representations which include
a $\hmm$ Higgs boson that has small (most naturally zero) $\wm\wm$ coupling 
but possibly non-zero $\ell^-\ell^-$ coupling. It is then
very possible that $B(\hmm\to\ell^-\ell^-)\sim 1$ for $\ell=e$, $\mu$, or 
(most probably?) $\tau$. We have demonstrated that 
detection of the $\hmm$ at the Tevatron
(operating at $\rts=2\tev$ with $L=30\fbi$) will then be possible for $\mhmm$ 
up to $300\gev$ for  $\ell=e$ or $\mu$ and  $180\gev$ for $\ell=\tau$.
We can estimate from Fig.~\ref{fig:xsec}
the corresponding limits at the LHC by requiring
the same raw number of events before cuts and efficiencies as
needed at the Tevatron --- $\sim 10$ for $\ell=e,\mu$ and $\sim 300$ for
$\ell=\tau$ --- yielding $\mhmm$ discovery up to roughly $925\gev$ ($1.1\tev$)
for $\ell=e,\mu$ and $475\gev$ ($600\gev$) for $\ell=\tau$,
assuming total integrated luminosity of $L=100\fbi$ ($L=300\fbi$).
For $\ell=e,\mu$, the reach of the 
LHC detectors will likely be even greater than
this, due to the improved lepton acceptance and resolution
anticipated over the current generation of hadron collider detectors.
For $\ell=\tau$, this simple extrapolation may not account
for a different signal-to-background ratio in 
$\tau$ selection at
the LHC.  A full study is necessary to evaluate this.

As detailed in \cite{ememgunion}, if a $\hmm$ is found
then $e^-e^-$ and $\mu^-\mu^-$ colliders capable of high luminosity
at $\rts=\mhmm$ will become a priority in order to actually determine 
the $c_{\ell\ell}$'s.
Indeed, observation of $\hmm\hpp$ pair production in only 
a single $\hmm\to \ell^-\ell^-$ channel provides
no information on $c_{\ell\ell}$. (Of course, if more than one $\ell\ell$
channel is seen, ratios of the $c_{\ell\ell}$'s could be obtained.)
Only if the $\hmm\rta\hm\wm$ decay channel [for which
the partial width can be computed and compared to the
$\ell^-\ell^-$ partial width via Eq.~(\ref{widtheq})]
is also seen, can one get an estimate of the $c_{\ell\ell}$ magnitude(s).
In contrast, an $e^-e^-$ ($\mu^-\mu^-$) collider would provide a 
direct measurement of $c_{ee}$ ($c_{\mu\mu}$).
For a more detailed discussion see \cite{ememgunion}. 
This illustrates an important complementarity between the NLC and 
hadron colliders. Discovery of a $\hmm$ prior to
the construction and operation of the $\epem,\emem$ collider NLC complex
would be very important in determining the energy range over which good 
luminosity and good energy resolution for $\emem$ collisions should be a
priority.

%

%

\end{document}